\documentclass[amstex]{article}
\usepackage{amssymb}

\usepackage{graphicx}
\usepackage{amsmath}

\input{epsf.sty}

\newtheorem{theorem}{Theorem}

\newtheorem{proposition}[theorem]{Proposition}

\input{tcilatex}

\begin{document}

\date{January 2002}


\thispagestyle{empty} \hfill{\footnotesize CBPF-NF-005/02}

\begin{center}
{\large \textbf{Area Law for Localization-Entropy in Local Quantum Physics}}

\vspace{0.2cm}

\textsl{Bert Schroer\\[0pt]
presently: CBPF, Rua Dr. Xavier Sigaud 150 \\[0pt]
22290-180 Rio de Janeiro - RJ, Brazil \ \\[0pt]
Prof. em. of the FU-Berlin, Germany\\[0pt]
email schroer@cbpf.br}


January 2002
\end{center}

\begin{abstract}
The previously developed algebraic lightfront holography is used in
conjunction with the tensor splitting of the chiral theory on the causal
horizon. In this way a universal area law for the entanglement entropy of
the vacuum relative to the split (tensor factorized) vacuum is obtained. The
universality of the area law is a result of the kinematical structure of the
properly defined lightfront degrees of freedom. We consider this entropy
associated with causal horizon of the wedge algebra in Minkowski spacetime
as an analog of the quantum Bekenstein black hole entropy similar to the way
in which the Unruh temperature for the wedge algebra may be viewed as an
analog in Minkowski spacetime of the Hawking thermal behavior. My more
recent preprint hep-th/20202085 presents other aspects of the same problem.
\end{abstract}

\vspace{0.5cm}

\hspace{0.2cm} \textbf{Key-words:} Quantum Field Theory; Mathematical
Physics; Thermal Physics

\newpage\setcounter{page}{1}

\section{Introductory remarks}

``Localization Entropy'' \cite{Narn}\cite{S1} is an entropy which has its
origin in the fact that in local quantum physics\footnote{%
This is the formulation of QFT without the use of field-coordinatizations.}
(in distinction to quantum mechanics) the vacuum, if restricted to a
causally closed algebra with a nontrivial causal complement, behaves as a
thermal state \cite{Haag}. The best known illustration is the vacuum
restricted to the algebra of quantum matter localized in a Rindler wedge
which leads to the planar Unruh situation i.e. a thermal state with a
Hawking temperature (which depends on the horizon-generating acceleration in
the Unruh Gedankenexperiment) which is accompanied by thermal radiation \cite
{Wald}. Whereas the thermal aspect of such a situation is related to the
general observation that spacetime restrictions of globally pure states in
QFT to regions with a nontrivial causal disjoint lead inevitably to impure
states (not true in QM), the specific understanding of an entropy generated
by ``quantum'' localization and its relation with the classical
Bekenstein-Hawking area behavior is a more challenging task \cite{S1}.

In this work we use the recently proposed algebraic lightfront holography 
\cite{S2} in order to derive a transverse tensor factorization structure of
degrees of freedom. This means that the previously obtained generalized
chiral operator algebra on the lightfront $\mathcal{A}(LF),$ whose main
distinction from a standard chiral algebra is a huge transverse degeneracy,
possesses a transverse tensor product foliation into nonoverlapping chiral
longitudinal strip algebras $\mathcal{A}(\mathcal{S}_{i})$ (tiling of $LF$
into strips of unit transverse width) which are also generalized chiral
conformal algebras

\begin{align}
\mathcal{A}(LF) & =\overline{\bigotimes_{i}}\mathcal{A}(\mathcal{S}_{i}) \\
H & =\overline{\bigotimes_{i}}H(\mathcal{S}_{i})  \notag
\end{align}
The restriction of this tiling to the causal horizon of a wedge $W$ (the
Unruh-Rindler situation) defines a tensor product tiling on the
half-lightfront $LF_{+}$ (the upper causal horizon of $W)$%
\begin{align}
\mathcal{A}(LF_{+}) & =\overline{\bigotimes_{i}}\mathcal{A}(\mathcal{S}%
_{+,i}) \\
H_{i} & =\overline{\mathcal{A}(\mathcal{S}_{+,i})\Omega}  \notag
\end{align}
In the next section we will review the derivation of the chiral nature of
the lightfront algebra and explain how from its additional transverse
structure one arrives at this factorization.

Quantities as the entropy (which behave additively under tensoring into
independent i.e noninteracting subalgebras) will then be naturally described
in terms of an entropy density per unit $d-2$ dimensional transverse cell
(taking the $d-2$ dimensional transverse instead of a naively expected d-1
dimensional horizontal density). However it is well-known among algebraic
quantum field theorist \cite{Haag} that the very nature of these algebras
(the $\mathcal{A}(\mathcal{S}_{+,i})$ are hyperfinite type III$_{1}$ von
Neumann factors) prevents a direct assignment of entropy. To be more
specific, although the vacuum tensor-factorizes in transversal direction
(i.e. there is no transverse vacuum polarization), a factorization in
lightray direction is not possible i.e. there is no isomorphism of operator
algebras 
\begin{equation}
\mathcal{A}(\mathcal{S})\nsim\mathcal{A}(\mathcal{S}_{-,i})\overline
{\bigotimes}\mathcal{A}(\mathcal{S}_{+,i})
\end{equation}
The physical reason are the uncontrollable vacuum fluctuation which the
``tearing apart'' of a type I global algebra $\mathcal{A}(\mathcal{S})$ into
two type III$_{1}$ algebras causes at the common boundary. To arrive at a
situation in which a longitudinal tensor factorization can be defined, we
must find a way to control the vacuum polarization with the help of a more
careful ``splitting procedure''. Buchholz and Wichmann have shown that each
QFT with a reasonable thermal behavior fulfills the split property \cite
{Haag}. Its adaptation to the strip algebras is the subject of the third
section in which it is shown that the original vacuum becomes highly
entangled in terms of a suitable tensor product split which leaves a small
finite lightlike separation $\delta$ between the two half-strips $\mathcal{S}%
_{\pm,i}.$ There the reader finds also a formula for the overlap between the
vacuum and its split version as a function of $\delta.$ The looked for
split- or localization-entropy per unit d-2 dim. transverse volume (area for
d=1+3) which measures the entanglement is the relative entropy between the
vacuum and its split version considered as states on the strip algebra $%
\mathcal{A}(\mathcal{S}_{+})\subset\mathcal{A}(LF_{+}).$ Its calculation is
reduced to computations within ``kinematical'' ((half)integer scale
dimensions)\ chiral theories, a fact which accounts for the insensitivity of
the result on the original quantum matter content. In the concluding remarks
we compare our results to previous attempts and mention some open problems.

Since the phenomenon of vacuum polarization (which separates quantum
mechanics from local quantum field theory) and in particular its behavior
under holographic lightfront projection is the main cause for transverse
alignment of degrees of freedom on the lightfront, we will use the remainder
of this introduction to recall some historical facts which help to
understand the setting of this paper.

Historically vacuum polarization was first observed by Heisenberg, Weisskopf
and others \cite{QED} while trying (in contemporary terminology) to define
global Noether charges as limits of ``partial charges'' in open systems.
Partial charges associated to sharply defined spatial regions diverge
because the vacuum fluctuations which such an object creates if applied to
the vacuum become uncontrollably big. Although the intuitive remedy in form
of a ``soft'' boundary was quite clear, careful mathematical definition in
terms of smearing functions for partial charges only appeared when (as a
result of the appearance of the notion of a spontaneously broken symmetry)
there was a need for a higher precision on this point \cite{Haag}\cite{Requ}%
. In a way the split property may be viewed as the algebraic version of this
taming of vacuum fluctuation. Although we have not found a direct connection
between Noether currents and the problem of localization-entropy (since this
entropy is a pure local quantum physics notion, it would be difficult to
imagine that it arises from quantization of a classical Noether current), it
may be interesting to point out that Wald in his earlier work within a more
classical gravity setting \cite{Wa} has given a description in terms of
Noether currents. In addition a Noether formalism is used in the third
section as an auxiliary device to implement a representation for the split
vacuum state. Whether these observations are coincidental or deeply
connected remains a problem of future research.

The fact that causality requires the presence of both frequencies, and in
this way causes vacuum polarization, is also the mechanism behind the
Reeh-Schlieder theorem. In the standard literature known under the name of \
``state-algebra relation'' (i.e. the unique relation between operators of a
local operator algebra and a dense set of vectors created by these operators
when they act on the vacuum), it contrasts local quantum physics from
quantum mechanics in such a dramatic way that it even attracted a lot of
attention among philosophers \cite{Hal}. The Reeh-Schlieder theorem is also
the starting point for the use of the Tomita-Takesaki modular theory for the
analysis of local algebras together with the vacuum (or other cyclic and
separating state vectors). Operator algebraist call a Reeh-Schlieder pair ($%
\mathcal{A},\Omega $) a ``standard'' pair. The first application of the
modular theory of operator algebras in QFT was elaborated by Bisognano and
Wichmann \cite{Bi-Wi} for the standard pair ($\mathcal{A}(W),\Omega$) where $%
W$ denotes a (Rindler) wedge. The relation between that theory and Unruh's
thermal observations about Rindler wedges (creation of a horizon by uniform
acceleration) was first seen in \cite{Sewell}. The content of the present
paper is best viewed as a refinement with the help of the lightfront
algebra. Although the lightfront $LF$ is the linear extension of the upper
causal horizon $h_{+}(W)=LF_{+}$ of the wedge $W$ and the two algebras have
been shown to be equal i.e. $\mathcal{A}(W)=\mathcal{A}(LF_{+}),$ the local
alignment of degrees of freedom in the $\mathcal{A}(LF_{+})$ description is
better suited to see the transverse factorization and to describe the
ensuing localization entropy caused by splitting the vacuum along the
bifurcation.

\section{Review of lightfront holography}

The algebraic lightfront holography \cite{S2} extends the old lightfront
quantization (or $p\rightarrow\infty$ frame method) to the realm of
interacting renormalizable QFT. Whereas the old method associates pointlike
fields on the lightfront by a suitably defined restriction of the free (or
superrenormalizable) fields, the lightfront holography reprocesses the
d-spacetime dimensional local algebras (generated by smeared pointlike
fields in case one starts from pointlike fields) of a wedge region\footnote{%
We need an algebra in ``standard position'' relative to the vacuum vector $%
\Omega$ i.e. one for which the algebra acts cyclic and has no annihilators
of $\Omega;$ hence we cannot take the full algebra on Minkowski space.
Whenever the localization region has a nontrivial spacelike complement as in
the case of a wedge, the ``standardness'' is guarantied.} (which is the
causal forward/backward shadow cast by a semi-lightfront) into a net of d-1
dimensional subalgebras of an algebra localized on the (upper) horizon (i.e.
the lightfront half-plane) of the right wedge. This algebraic lightfront
holography is applicable to all local field theories, whereas the old
lightfront approach only works for theories in which the integral over the
Kallen-Lehmann spectral function is finite (which is the same severe
restrictions as the one required by the validity of the canonical
formalism), which leaves only the free\footnote{%
In d=1+1 there are also superrenormalizable interacting models
(e.g.polynomially coupled scalar fields) which are canonical, but they are
not very interesting for particle physics.} models in d=1+3 \cite{S2}. The
algebraic approach simply liberates the causal propagation picture from its
narrow canonical limitation in terms of lightfront-restricted pointlike
fields.

Although this reprocessing maintains the longitudinal localization (in the
plane spanned by the two generating lightlike defining vectors of the
wedge), one looses the localization in the transversal direction. Of course
one may get this information by considering the intersection of all double
cone algebras which touch the lightfront, but since regions on the
lightfront with finite transversal/longitudinal extension do not cast a
``causal shadow'', one has to study the properties of d-dim. regions which
become infinitely thin in the direction perpendicular to the lightfront,
which makes the construction questionable. A better way, more in the spirit
of algebraic QFT is to recover the transversal net structure by
``Lorenz-tilting'' the wedge around its upper defining lightlike vector i.e.
the tilting belongs to the Wigner little group of that lightlike vector. By
forming algebraic intersections between the original and the transformed
strips in transversal direction one obtains a net structure on the
lightfront \cite{S2}.

The mathematical basis of the operator-algebraic holography is a theorem on
modular inclusion \cite{Wies} of two von Neumann algebras $\mathcal{N}\subset%
\mathcal{M}$ with a common cyclic and separating vector $\Omega$ such that
the modular group $\sigma_{t,\mathcal{M}\text{ }}$ of the pair ($\mathcal{M}%
,\Omega$) upon restriction to $\mathcal{N}$ compresses i.e. $\sigma_{t,%
\mathcal{M}\text{ }}(\mathcal{N})\subset\mathcal{N},\,t<0$ (in this case it
was called a +halfsided modular inclusion in \cite{Wies}). Modular
inclusions for which the relative commutant $\mathcal{N}^{\prime}\cap%
\mathcal{M}$ is also standard with respect to $\Omega$ are isomorphic to
chiral conformal theories \cite{Wies} and hence the transversely unresolved
lightfront holographic projection is a chiral theory. In fact it is a
``kinematical'' chiral theory in the sense that its pointlike generating
fields (which have no direct relation to the original fields) have
(half)integer scale dimensions.

For the case at hand $\mathcal{M}=\mathcal{A}(W),$ $\mathcal{N}=\mathcal{A}%
(W_{e_{+}})=AdU(e_{+})\mathcal{N}$ with $e_{+}=(1,1,0,0).$ We collect the
two most important results of these assumptions

\begin{itemize}
\item  The wedge algebra is equal to its upper horizon (lightfront
half-plane) algebra \cite{S2} 
\begin{equation}
\mathcal{A}(W)=\mathcal{A}(R_{+})  \label{char}
\end{equation}
It is enough to know the operator algebra for the standard $x$-$t$ wedge $W$
because the full net of local algebras may then be obtained by covariance
and intersections of algebras. The family of subalgebras $\mathcal{A}(\left[
a,\infty\right] )$ of $\mathcal{A}(R_{+})$ which is indexed by semi-infinite
intervals $\left[ a,\infty\right] \subset R_{+}$ is isomorphic to the family 
$\left\{ \mathcal{A}(W_{ae_{+}})\right\} _{a>0}$ whereas the original double
cone subalgebras $\mathcal{A}(W_{ae_{+}-be_{-}})$ $a,b>0$ have a fuzzy image
in $\mathcal{A}(R_{+})$ (i.e. no geometrically characterizable position in $%
\mathcal{A}(R_{+})$). Vice versa the interval-localized algebras $\mathcal{A}%
(I),I\subset R_{+}$ are holographic images of fuzzy localized subalgebras in 
$\mathcal{A}(W).$ The holographic relation can be extended to the full
algebra $\mathcal{A}$ which then becomes holographically encoded into the
full lightfront plane algebra $\mathcal{A}(R)=B(H)$

\item  The diffeomorphism group of the chiral $\mathcal{A}(R)=$ $\mathcal{A}%
(S^{1})$ is that of the circle (its infinitesimal generators obey the
Virasoro-algebra commutation relations), whose modular origin has been
recently established \cite{Lucio}. It has a holographic pullback to fuzzy
acting automorphisms on the original algebra in $d$ spacetime dimensions.
The subgroup of the circular diffeomorphism group which acts in a local
manner both on the original and the holographically projected algebra is the
group generated by dilation and lightray translation into the $e_{+}$
direction. The rigid circular rotation generated by Virasoro's $L_{0}$
belongs to the symmetries which act fuzzy on the original algebra. Those
Poincar\'{e} transformations which act locally on the lightfront form a
7-parametric subgroup.
\end{itemize}

Some more comments are in order.

Although the original as well as the projected theory may have a
conventional description in terms of pointlike field generators, the
algebraic holographic reprocessing of degrees of freedom in the presence of
interactions cannot be formulated in terms of field coordinates, but rather
needs the concepts of the operator-algebraic approach to QFT. The reason is
that these holographic maps involve steps which change the spacetime
indexing of operators in such a way that a geometrically localized algebra
goes into one with a ``fuzzy'' localization and vice versa, i.e. a geometric
localization in the holographic image may come from a fuzzy localized
subalgebra of the original theory \cite{S2}. The physical intuitive content
(but not the conceptual framework) of the present approach is close to 't
Hooft's area-law inspired holography \cite{'t Hooft}.

In order to have an easy geometric visualization of conformal theories
associated with lightfronts, we sketch the important step of the modular
inclusion method for d=1+2. Let $W$ be the standard x-t wedge and consider
the inclusion ($Ad$ denotes the adjoint action) 
\begin{align}
\mathcal{A}(W_{e_{+}}) & \subset\mathcal{A}(W) \\
\mathcal{A}(W_{e_{+}}) & \equiv AdU(e_{+})\mathcal{A}(W)  \notag \\
e_{+} & =\frac{1}{\sqrt{2}}(1,1,0)  \notag
\end{align}
The modular group $\sigma_{t}$ of $\mathcal{A}(W)$ is implemented by the $W$%
-fixing x-t Lorentz boost $\Lambda_{x-t}(-2\pi t);$ it evidently compresses $%
W_{e_{+}}$ for $t<0$ which is the defining property of a modular inclusion.
The relative commutant $\mathcal{A}(W_{e_{+}})^{\prime}\cap\mathcal{A}(W)$
of the shifted algebra $\mathcal{A}(W_{e_{+}})$ in $\mathcal{A}(W)$ is the
building block of the transversely unresolved lightfront algebra $\mathcal{A}%
(R_{+})$%
\begin{align}
\mathcal{A}(R_{+}) & \equiv\bigvee_{t}\sigma_{t}(\mathcal{A}%
(W_{e_{+}})^{\prime}\cap\mathcal{A}(W)) \\
\mathcal{A}(R) & \equiv\mathcal{A}(R_{+})\vee AdJ\mathcal{A}(R_{+})  \notag
\end{align}
Of course one must argue that the nontriviality of the relative commutant is
not tied to the existence of a generating field with canonical commutation
relations (decreasing Lehmann-Kallen spectral function) and that the
relative commutant has a cyclic action on the vacuum. For this we refer to
previous works \cite{S2}. The upshot is the relation (\ref{char}) , which is
the quantum analog of the classical statement that (with the exception of
conformal covariant d=1+1 theories) the characteristic lightfront data on
the upper (lower) wedge horizon determine the data within the wedge.

The gain of using the holographic projection onto the lightfront is that the
transversely unresolved lightfront algebra turns out to be a \textit{chiral
theory with additional automorphisms acting on it.} The simplicity of our
d=1+2 illustration lies in the fact that the transversal symmetries which
act as automorphisms of the lightfront algebra are easily recognizable. They
consist of the y-translation and the one-dimensional Wigner little group%
\footnote{%
It consists of a t-y Lorentz boost which would turn the original edge of the
wedge (of which the half lightfront is the upper horizon) outside of the
lightfront combined with an appropriately chosen x-y rotation which turns
the edge back into the lightfront but into a tilted position with respect to
its original location.
\par
{}} of the lightlike vector $e_{+}$ 
\begin{align}
x_{+} & \rightarrow x_{+},\text{\ }y\rightarrow y+vx_{+} \\
x_{-} & \rightarrow x_{-}+2vy+v^{2}x_{+}  \notag
\end{align}
which acts as a kind of transversal Galilei transformation in the lightfront
plane 
\begin{align}
y & \rightarrow y-vx_{+}  \label{G} \\
x_{+} & \rightarrow x_{+}  \notag
\end{align}
Hence the total symmetry group acting on the 2-dim. lightfront plane is a
4-parametric subgroup of the 6-parametric Poincar\'{e} group (in d=1+3 the
lightfront symmetry group is 7-dimensional and there are 2 Wigner
``translations'' whose holographic projection looks like a transversal
Galilei transformation in the 3-dim. lightfront plane.) The two remaining
transformations which lead out of the 2-dim. lightfront plane are the
spatial rotation and the lightlike translation perpendicular to the plane.
The transformation (\ref{G}) may be used to equip the lightfront with a net
structure by intersecting the transverse strips corresponding to a
longitudinal (lightlike) intervals with their images under (\ref{G}) which
are inclined strips. Our main interest is in the quantum physical behavior
of longitudinal strips which may be obtained from the net by additivity of
the lightfront net. Let us imagine that we have divided the lightfront plane
into horizontal strips $S_{i}$ $i\in\mathbb{Z}$ of unit transversal width.
These strips have the special property of containing just one lightlike
direction, all other directions are spacelike. This together with the fact
that the lightlike generator is positive has the following well-known
consequence

\begin{proposition}
(\cite{Driessler}) The weakly closed algebras $\mathcal{A}(S_{i})$ localized
in the strips are type I$_{\infty}$ tensor factors whose tensor product is
the (weak closure of the) global algebra 
\begin{equation}
\bigotimes_{i}\mathcal{A}(S_{i})=B(H)
\end{equation}
The transversal causal decoupling is ``quantum mechanical'' i.e. in the
sense of transversal factorization of the vacuum 
\begin{equation*}
\Omega=\prod_{i}\otimes\Omega_{i}
\end{equation*}
\end{proposition}

The crucial property used in the proof is the existence of a positive
generator lightlike translation $U_{x_{+}}(a)$ which leaves each $\mathcal{A}%
(S_{i})$ invariant. This leads to 
\begin{equation}
\left\langle AdU_{x_{+}}(a)A_{i}\cdot A_{j}\right\rangle =\left\langle
A_{j}\cdot AdU_{x_{+}}(a)A_{i}\right\rangle ,\,\,i\neq j
\end{equation}
which together with the analytic properties in $a$ following from the
positivity of the lightlike momentum leads to the independence on $a$ which
in turn combined with the cluster property yields the vacuum factorization
property. This property also holds for the half strips which partition the
half lightfront (the horizon of $W)$ since the associated algebras are
subalgebras of the $S_{i}.$

This shows a somewhat unexpected (complete absence of transverse vacuum
fluctuations) behavior know from a collection of quantum mechanical systems
without relative interactions between them. Since the lightfront degrees of
freedom factorize in this perfect way we encounter an ``area law'' in the
sense that if we would be able to measure the degrees of freedom in a strip
in terms of a strip entropy, then this entropy would have the interpretation
of the entropy per unit edge size of the edge of the Unruh-Rindler-wedge $W.$

It has been noted previously \cite{S2} that those local Poincar\'{e}
automorphisms which lead out of the lightfront become fuzzy (non-geometric)
under holographic projection. In the other direction there are
transformations which in the sense of the lightfront net structure are local
and which if reprocessed into the higher dimensional geometric setting will
be fuzzy: an example is the conformal rotation (its generator is often
referred to as the ``conformal Hamiltonian $L_{0}$). As a result of the
modular origin\footnote{%
The modular origin substitutes the existence of the energy-momentum tensor.
As a result of the transversal structure the strip algebras are much larger
chiral algebras than those standard chiral algebras resulting from d=1+1
conformal QFT.} of the diffeomorphisms of the circle \cite{Lucio}\cite{Fard}%
, this diffeomorphism group also acts locally on the strip algebras and
lifts to fuzzy symmetries in the original higher dimensional setting. For
this reason the lightfront holography represents a more radical reprocessing
than Rehren's AdS-CQFT isomorphism \cite{Rehren} (which is the rigorous
field-theoretic version of Maldacena's conjecture).

Clearly there is nothing in our d=1+2 sketch which has no counterpart in d%
\TEXTsymbol{>}1+2 spacetime dimensions. The transverse transformations in
the general case consist of ordinary transverse translations and
translations within the Wigner little group of $e_{+}$ which is the
Euclidean group in d-2 dimensions (containing d-2 transverse ``Galilei''
transformations). With the help of the latter one can resolve the
transversal net structure \cite{S2}.

\newpage

\section{The split isomorphism and its implementations}

Since the chiral strip algebras restricted to the horizon of $W$ are not
ordinary quantum mechanical algebras (i.e. type I von Neumann algebras with
minimal projectors associated with pure states corresponding to best
possible measurements), a direct attempt to associate to it a von Neumann
entropy\footnote{%
Even if the absolute entropy is infinite, the relative entropy between two
different states on the same algebra can be finite.
\par
{}} is meaningless. As already indicated in the last part of the
introduction, the non quantum mechanical nature of local operator algebra
(hyperfinite von Neumann factor of type III$_{1})$ does not allow the
definition of a von Neumann entropy; in fact such operator algebras do not
even permit tracial states/weights. Since there are no pure states, the
notion of entanglement is also void of meaning.

The remedy of passing to localized algebra with a fuzzy boundaries of
arbitrary small but finite thickness by splitting was already mentioned in
the introduction \cite{D-L}\cite{Haag}. Let us present it first in a more
general setting before adapting it to the kind of chiral theory which
represent the strip algebras.

We start from an inclusion of two double cones $\mathcal{C}_{1},\mathcal{C}%
_{2}$, but different from the physical realizations of modular inclusion
where the causal horizons touch each other, the causal boundaries of a split
inclusion are separated by a ``collar'' of thickness $\delta$%
\begin{equation*}
\mathcal{C}_{1}\subset\subset_{\delta}\mathcal{C}_{2}
\end{equation*}
In theories fulfilling the Buchholz-Wichmann nuclearity (a phase space
property which insures a reasonable thermal behavior) such split inclusions 
\cite{Haag} lead to the existence of an intermediate type I factor which is
contained in the bigger and contains the smaller operator algebra, but has
no sharp boundaries within the collar. This existence of an intermediate
type I factor (without geometric interpretation) constitutes the definition
of a split inclusion whose mathematical aspects have been investigated in
detail in \cite{D-L}. In fact there it was shown that if the inclusion $%
\mathcal{A}(\mathcal{C}_{1})\subset\mathcal{B}(\mathcal{C}_{2})$ splits in
this fashion, there exists even a distinguished intermediate canonical type
I factor $\mathcal{N}$ functorially related to $\mathcal{A(C}_{1}\mathcal{)},%
\mathcal{B(C}_{2}\mathcal{)}.$ Suppose now that the reference state $\omega$
is also faithful on the algebra $\mathcal{A(C}_{1}\mathcal{)}\vee \mathcal{%
B(C}_{2}\mathcal{)}^{\prime}$ i.e. the operator algebra generated by the
algebra $\mathcal{A(C}_{1}\mathcal{)}$ and the commutant of $\mathcal{B(C}%
_{2}\mathcal{)}$. Such a state retains this property upon restriction to the
generating subalgebras and it is easy to prove that if one forms the product
state $\omega\cdot\omega$ by eliminating the correlations between these two
subalgebras, this new product state is implemented by a vector $\eta$ in the
common Hilbert space (by using an appropriate natural cone representation, $%
\eta$ will even be unique) \cite{Haag} 
\begin{align}
& \omega\cdot\omega(AB^{\prime})\equiv\omega(A)\omega(B^{\prime}),\,\,A\in%
\mathcal{A(C}_{1}\mathcal{)}\subset\mathcal{B(C}_{2}\mathcal{)}%
,\,B^{\prime}\in\mathcal{B(C}_{2}\mathcal{)}^{\prime} \\
& \,\curvearrowright\exists\,\eta\in H\text{ \thinspace}s.t.\,\omega
(\cdot)=\left\langle \eta\left| \cdot\right| \eta\right\rangle ,\,\,\;%
\mathcal{A(C}_{1}\mathcal{)\subset N\subset B(C}_{2}\mathcal{)}  \notag \\
& B(H)\simeq\mathcal{N}\bar{\otimes}\mathcal{N}^{\prime},\text{ }H=H_{%
\mathcal{N}}\overline{\otimes}H_{\mathcal{N}^{\prime}},\,H_{\mathcal{N}%
}\equiv\overline{\mathcal{A}(\mathcal{C}_{1})\eta}=P_{\mathcal{N}}H  \notag
\\
& \mathcal{N}=P_{\mathcal{N}}B(H)P_{\mathcal{N}},\text{ }\mathcal{A}(%
\mathcal{C}_{1})\subset\mathcal{N},\,\mathcal{B}(\mathcal{C}_{2})^{\prime
}\subset\mathcal{N}^{\prime}  \notag
\end{align}

Hence the desired split is accomplished by the type I algebra\footnote{%
There exists a concrete formula for the type I factor $\mathcal{N}$ in terms
of $\mathcal{A}(\mathcal{C}_{1})$ and the modular involution $J_{collar}$ of
the collar algebra $\mathcal{A}(\mathcal{C}_{1})^{\prime}\cap\mathcal{B}(%
\mathcal{C}_{2})$ \cite{D-L}. Physical intuition suggests that the
dominating behavior in the limit of small collar size is independent of the
chosen type I interpolation.} $\mathcal{N}$ and the state vector $\eta$ is a
tensor product vector without entanglement in the tensor product description 
$\mathcal{N}\otimes\mathcal{N}^{\prime}$ and hence a fortiori on $\mathcal{A(%
}\mathcal{C}_{1}\mathcal{)}\otimes\mathcal{B(}\mathcal{C}_{2}\mathcal{)}%
^{\prime}\subset\mathcal{N}\otimes\mathcal{N}^{\prime}.\,$On the other hand
the original non-split vacuum is highly entangled with the consequence that
it is not only a thermal state (with the Hawking temperature) on $\mathcal{A(%
}\mathcal{C}_{1}\mathcal{)}\otimes\mathcal{B(}\mathcal{C}_{2}\mathcal{)}%
^{\prime},$ but even remains thermal on the type $I$ quantum mechanical
algebra $\mathcal{N}$. The localization of this ``relativistic box
quantization'', contrary to a quantum mechanical box, is fuzzy inside the
collar $\mathcal{C}_{1}$\TEXTsymbol{\backslash}$\mathcal{C}_{2}.$ But it
needs to be emphasized that contrary to the inside/outside nonrelativistic
box quantization no degrees of freedom have been dumped or cut-off in the
present case; they were only somewhat spatially displaced within the collar
region in order to avoid uncontrollable vacuum fluctuations from the sharp
splitting.

For the case at hand the necessary splitting is in a \textit{generalized}
chiral theory$.$ Here ``generalized'' means that there is a large transverse
multiplicity which is encoded into the finite width of the strip $S=\dot
{R}\times width,$ whereas standard chiral theories, which result from the
tensor-product decomposition of 2-dim. conformal models, are depicted as
being localized on the compactified line $\dot{R}\simeq S^{1}$. The
splitting into two $S_{\pm,\delta}=R_{\pm,\delta}\times width$ with $%
S_{\pm,\delta}\subset LF_{\pm},$ where $S_{\pm,\delta}$ correspond to the
above $\mathcal{A(}\mathcal{C}_{1}\mathcal{)},\mathcal{\mathcal{B(}\mathcal{C%
}}_{2}\mathcal{\mathcal{)}^{\prime}},$ is done by splitting $\dot{R}\simeq
S^{1}$ symmetrically at lightlike zero and infinity by a small $\delta;$ the
two small disconnected splitting intervals correspond to the above collar.
For the explicit description of the split vacuum $\eta$ it is important to
notice that the chiral theories which originate through lightfront
holography are ``kinematical'' in the sense that they only contain canonical
((half)integer) scale dimensions. In fact such chiral algebras always
possess pointlike field generators of (half)integer scale dimensions \cite
{Joerss}.

Let us first look at a standard chiral situation i.e. let us ignore the
transverse extension. For simplicity assume that the generating field has
scale dimension $d=\frac{1}{2}$

Let us furthermore consider the special case that the chiral theory has a
generating field of scale dimension $d=\frac{1}{2}.$ The two $\delta$-split
even operator algebras are then of the simple form

\begin{equation}
\mathcal{A}(\left( R_{\pm,\delta}\right) )=alg\left\{ \psi(f)\psi^{\ast
}(g)|\,suppf,g\in R_{\pm,\delta}\right\}
\end{equation}
and the split isomorphism $\Phi$ acts as 
\begin{equation}
\Phi\left( \mathcal{A}(R_{-,\delta})\vee\mathcal{A(}R_{+,\delta}\right) )=%
\mathcal{A}(R_{-,\delta})\bar{\otimes}\mathcal{A}(R_{+,\delta})
\end{equation}
We are interested in the relation of the original vacuum $\Omega$ to the
split vacuum $\eta$ 
\begin{align}
& \eta\simeq\Omega\bar{\otimes}\Omega  \label{prod} \\
& \left\langle \eta\left| \mathcal{A}(R_{-,\delta})\vee\mathcal{A(}%
R_{+,\delta}\right| \eta\right\rangle =  \notag \\
& \left\langle \Omega\left| \mathcal{A}(R_{-,\delta})\right| \Omega
\right\rangle \cdot\left\langle \Omega\left| \mathcal{A(}R_{+,\delta}\right|
\Omega\right\rangle  \notag
\end{align}
in the limit of shrinking split size $\delta$. The choice of the
implementing vector $\eta$ is not unique; a mathematically preferred choice
is obtained by taking $\eta$ in the natural cone of the standard pair \cite
{D-L}\cite{Haag} $\mathcal{A}(\left( -\infty+a,-a\right) )\vee\mathcal{A(}%
(a,\infty -a)),\Omega).$ Any other choice $\eta^{\prime}$ is related to the
canonically preferred $\eta$ by the following inequality \cite{charges} 
\begin{align}
\left\| \eta-\Omega\right\| ^{2} & =2\left| 1-\left( \eta,\Omega\right)
\right| \\
& =inf\left\{ \left\| \eta^{\prime}-\Omega\right\| ^{2},\eta^{\prime
}\,is\,\,split\right\}  \notag \\
& =\left\| \omega_{split}-\omega_{\Omega}\right\| \leq\left\| \eta
^{\prime}-\Omega\right\| ^{2}  \notag
\end{align}
where the last line uses the so-called canonical Bures distance in the
convex space of states. Bures distance 2 is an indication that the state $%
\omega_{split}$ belongs to an inequivalent folium (its GNS representation
defines an inequivalent representation of the chiral algebra). It is
generally believed that in the limit $\delta\rightarrow0$ all implementing
state vectors show the same behavior. We will assume that this is true, and
that there is no physically preferred implementation.

The main point of this section is now the proposition that thanks to the
simple kinematical structure of the holographic projection, the ``escape''
into an inequivalent representation in the limit $\delta\rightarrow0$ \cite
{Wald} as well as the entanglement of the vacuum and the resulting
localization-entropy (and its divergence with shrinking collar size) can be
studied quantitatively with the help of the ``flip trick'' which is an
implementation of $\Phi$ in terms of concrete unitary operators. For the
case at hand we notice that the fields in the different tensor product
factors can be interpreted as a doublet i.e. $\psi_{1}=\psi\bar{\otimes}%
1,\,\psi_{2}=1\bar{\otimes}$ $\psi.$ In the spirit of a SO(2) Noether
symmetry the implementation of the unitary flip operation can then be done
in terms of a Noether current \cite{charges} formalism 
\begin{align}
\Phi(\psi(x)) & =e^{ij(f)}\psi_{2}(x)e^{-ij(f)}=\left\{ 
\begin{array}{c}
\psi_{1}(x),\,x\in R_{+,\delta} \\ 
\psi_{2}(x),\,x\in R_{-,\delta}
\end{array}
\right. \\
j(x) & =\psi_{2}^{\ast}(x)\psi_{1}(x),\,\,\,f=\left\{ 
\begin{array}{c}
1,\,\,\text{\thinspace}x\in R_{-,\delta} \\ 
0,\,\,\,\,x\in R_{+,\delta}
\end{array}
\right.  \notag
\end{align}
Clearly $U(f)=e^{ij(f)}$ acting on $H\bar{\otimes}H$ implements the product
state (\ref{prod}) for $\left( \mathcal{A}(R_{-,\delta})\vee\mathcal{A(}%
R_{+,\delta}\right) .$ As expected the state vector $\eta^{\prime
}=U(f)(\Omega\bar{\otimes}\Omega)$ becomes orthogonal on all vectors in $H%
\bar{\otimes}H$ for $\delta\rightarrow0$ $.$ Let us check this for the
vacuum $\Omega\bar{\otimes}\Omega=\Omega_{vac}$%
\begin{align}
& \left\langle \Omega_{vac}|\eta^{\prime}\right\rangle =\left\langle
\Omega_{vac}\left| U(f)\right| \Omega_{vac}\right\rangle  \label{over} \\
& =e^{-\frac{1}{2}\left\langle j(f),j(f)\right\rangle _{0}}\sim
0,\,\,\delta\rightarrow0  \notag \\
& \left\langle \eta\left| AB\right| \eta\right\rangle =\left\langle
\Omega_{vac}\left| A\right| \Omega_{vac}\right\rangle \left\langle
\Omega_{vac}\left| B\right| \Omega_{vac}\right\rangle  \notag \\
& A\in\mathcal{A(}R_{-,\delta}\mathcal{)},\,\,B\in\mathcal{A(}R_{+,\delta }%
\mathcal{)}  \notag
\end{align}
The norm square of the smeared current $j(f)$ can be explicitly computed
from the known current two-point function 
\begin{equation*}
\left\langle j(f),j(f)\right\rangle _{0}\overset{\delta\rightarrow0}{\sim }%
-\log\delta
\end{equation*}
i.e. the resulting logarithmic dependence on the collar size $\delta$ leads
to a positive power law for the after exponentiation. This is a quantitative
expression for Wald's qualitative discussion of the inequivalence
(orthogonality) of the two Hilbert spaces \cite{Wald}.

In the same vein the inner product with all basis vectors converges with a
power law to zero for $\delta\rightarrow0$%
\begin{equation}
\left\langle \Omega_{vac}\left| U(f)\right|
a_{1}^{\ast}(p_{1})...a_{n}^{\ast}(p_{n})\otimes
a_{2}^{\ast}(k_{1})...a_{2}^{\ast}(k_{m})\Omega_{vac}\right\rangle
\rightarrow0
\end{equation}
i.e. the original vacuum becomes a highly entangled state on the split
algebra which in the limit $\delta\rightarrow0$ even leaves the Hilbert
space i.e. belongs to an inequivalent representation of the algebra. By
tracing out the first tensor factor one expects to obtains a density matrix
in the second factor which represents the vacuum $\Omega_{vac}$ as a mixed
state on the factor space $\overline{\mathcal{A(}R_{+,\delta}\mathcal{)\eta}%
^{\prime}}.$ In this cumbersome way one could try to compute the split
entropy, but it would be difficult to see more than its logarithmic
divergence which corresponds to power law which describes the approach to
the inequivalent representations. Fortunately there is a neater way due to
Kosaki \cite{Kosaki} which presents this relative entanglement entropy of $%
\omega$ and $\omega^{2}\equiv \omega\cdot\omega$ in terms of a variational
problem 
\begin{align}
& S(\omega^{2},\omega)= \\
& sup_{y(t)}\int_{0}^{\infty}\left[ \frac{\omega^{2}(1)}{1+t}-\omega
^{2}(y^{\ast}(t)y(t))-\frac{1}{t}\omega(x(t)x^{\ast}(t))\right] \frac{dt}{t}
\notag \\
& x(t)\equiv1-y(t)  \notag
\end{align}
here $y(t)$ is a path in the algebra $\mathcal{A}(R_{-,\delta})\vee \mathcal{%
A}(R_{+,\delta})$ which in our model is a Weyl algebra$.$ Note that this
variational formula is completely independent of which vector realization $%
\eta^{\prime}$ one chooses for the product state $\omega^{2}.$

Such formulas have been used for estimates of split entropies in \cite{Narn}%
. We hope to be able to return to a more direct calculation along the
indicated line.

Our strip algebras are different from standard chiral algebras in that we
are dealing with a collection of $\psi^{\prime}s$ indexed by points on the
transversal interval. But this only means that the flip current involves an
additional integration over a transverse unit cell which does not
participate in its commutation relations with the $\psi.$ Hence the
auxiliary Noether trick is still applicable.

Even after having understood the insensitivity of the dependence of quantum
area law on the concrete matter content in terms of the universality of the
lightfront holography, there remains the problem of the logarithmic
divergence with shrinking collar size which the classical Bekenstein formula
does not show. In fact as shown in \cite{Wa} the classical discussion
relates the entropy with a classical Noether current without vacuum
polarization effects. However in local quantum physics the dependence of
quantum entropy on the spatial distribution (localization) of degrees of
freedom and not only on their total number is an unavoidable consequence of
local quantum physics (vacuum polarization). In fact it is inexorably linked
to the transverse area behavior and it will not going away in curved
spacetime as the result of presence of curvature. However if the
localization entropy enters thermodynamic laws as the usual heat bath
entropy does, one could expect that the other quantum matter dependent terms
in such a law also show this vacuum polarization effect. In such a case it
is conceivable that this $\delta $-dependence could be scaled out of the
equation.

\section{Concluding remarks}

In this note we have proposed concepts which explain the universality of a
area law for localization quantum entropy in terms of the kinematical nature
of holographically projected degrees of freedom. This was achieved by
encoding as much as possible of the more complicated aspects into the
actions of automorphisms in order to keep the maximal simplicity of the
substrate (i.e. the lightfront) on which these automorphisms act.

Although this kinematical simplification aspect is very much in line with
why at the beginning of the 70s particle physicist became interested in the
use of lightfront and $p\rightarrow\infty$ methods, the present formulation
uses quite different concepts. The underlying strategy is to \textit{%
abstract from free fields only those properties which do not depend on short
distance property}.

This is why we had to reject the traditional restriction of pointlike fields
to the lightfront; it suffers from the same short distance limitation as the
canonical equal time formalism i.e. in d=1+3 theories all interactions would
be excluded; even asymptotically free theories (or supersymmetric theories)
do not allow a lightfront restriction since the asymptotic freedom property
does not make the integral over the spectral Kallen-Lehmann function
convergent. If on the other hand we generalize the \textit{modular aspects}
of operator algebras generated by free field to the realm of interactions
there is no such problem. However this strategy which avoids
field-coordinatizations has its price in that one looses the transverse
localization properties. Fortunately the 7-parametric symmetry group of the
lightfront algebra contains transformations from the Wigner little group of
the unique lightray in the lightfront which allow to recreate the net
structure of localized algebras on the lightfront, so that its transverse
tensor product foliation can be studied. We believe that there is much more
to these observations in that the structure of the holographic lightfront
projection may turn out to be the starting point of a new constructive
approach to QFT.

The new aspects of our calculation become more transparent if one compares
with the approach of Bombelli et al. \cite{Sor}. These authors do explicit
entropy computations on a box-localized (in the spirit of nonrelativistic
box quantization) zero mass free field, assuming that entropy is well
defined. They then find that it isn't, which forces them to cutoff
integrals. As far as the conservative field theoretic setting of these
authors is concerned, the calculations have a lot in common. The main
difference is that in the present approach no degrees of freedom have been
thrown away by cutting off integrals or in any other way. The split
inclusion method is a subtle but natural method which reprocesses the
original situation into one in which the inside/outside causal factorization
of degrees of freedom can take place.

An important step is to detach the degree of freedom issue from ``field
coordinatization'' (and their associated short distance properties) and tie
it directly to the structure of the restriction of the vacuum to the local
algebra. This very step already removes one source through which ultraviolet
divergencies usually appear, namely the short distance properties of the
(singular) field-coordinatization. Without focussing on algebras (and their
degrees of freedom) and shifting the emphasis away from field coordinates,
it would not be possible to consider the vacuum as an entangled state vector
with respect to a causal inside/outside division. Even if the original
theory and its holographic projection are both generated by pointlike
fields, in the presence of interactions there is still no direct relation
between them; our algebraic holographic method avoids being lured into such
wrong ideas where short distance conformal theories may get confused with
conformal aspects of lightfront holography. Although the holographic
projection is a conformal theory, it is a kinematical chiral theory which
(in the presence of interactions) has nothing to do with the scaling limit
of the original theory \cite{S2}. \textit{The distinction between short
distance universality classes and the present kinematical aspects of
lightfront holography is an important issue}.

It would be important to test the transverse tensor factorization in
rotational symmetric situations i.e. for the lightcone horizon of double
cone algebras. Here the difficulty is the absence of a global Killing
symmetry in the characterization of the causal horizon. Nevertheless there
are interesting analogies. For this and other problems which were omitted in
the present work, we refer to \cite{trans}. There the reader also finds more
details concerning the mathematical aspects of the lightfront QFT.

The main effect of curved spacetime seems to be a ``geometrization'' of
modular properties of operator algebras. Whereas in Minkowski spacetime the
Rindler restriction is the only way to have a causal horizon in terms of a
global Killing symmetry, curved spacetime creates such situations (together
with singularities) even in case of compact regions of bifurcation as in the
case of the Schwarzschild spacetime. If one views QFT as an abstract functor 
\cite{BFV} between spacetimes and algebras, then the different spacetimes
only give different localization textures to the same algebraic substrate
(e.g. the degrees of freedom of the Weyl algebra) and the properties of
localization entropy for causal horizons with Killing symmetries become more
visible on the classical side.

The kinematic chiral theory of the lightfront holography possesses the
Virasoro structure and as we have demonstrated recently, the diffeomorphism
of the circle have a modular origin and correspond to symmetries in the
common Hilbert space which act in a fuzzy way on the original non chiral
degrees of freedom before they suffered the holographic projection \cite
{Lucio}. This shows that the chiral structures used by Carlip \cite{Carlip}
in his entropy discussion are available without making additional
assumptions. However the identification of entropy with that of the
temperature state of the global chiral $L_{0}$ circular rotation
``Hamiltonian'' (in order to be able to apply the Cardy formalism) appears
somewhat ad hoc\footnote{%
According to Rehren \cite{Rehren} the rotational conformal operator becomes
the bona fide Hamiltonian if one re-processes the spacetime labeling of a
conformal theory into a AdS description which has a compact time coordinate.}%
. If the entanglement entropy of the chiral theory in the present treatment
would be equal to the Cardy entropy obtained from rotational $L_{0}$ thermal
states, then this would appear like an accident on the present level of
understanding. To clarify this, one needs more detailed investigations. In
any case the present explanation in terms of relative entanglement, unlike
most other attempts, require no additional degrees of freedom than those one
is dealing with in QFT in CST.

If there is any message about ``Quantum Gravity'' at all in the present
approach, perhaps it should be looked for in a better understanding of
possible relations between geometry and thermal behavior mediated by modular
theory rather than to a quantization sub prima facie of classical general
relativity. Although this message may sound pretty wild, recent results on
the construction of external and internal symmetries and spacetime geometry
from the relative position of operator algebras and in particular the
emergence of infinite dimensional fuzzy analogs of diffeomorphism groups
(including the Poincar\'{e} and conformal diffeomorphisms) from modular
inclusions and intersections of algebras point into the same direction \cite
{Wies}\cite{Bu}\cite{S-W}\cite{S1}.

\textbf{Acknowledgement}: I am indebted to Stefan Hollands and T. Thiemann
for suggesting some helpful references


\begin{thebibliography}{99}
\bibitem{Narn}  H. Narnhofer, in ``\textit{The State of Matter}'', ed. by M.
Aizenman and H. Araki (World-Scientific, Singapore) 1994

\bibitem{S1}  B. Schroer, J. Math. Phys. \textbf{41}, (2000) 3801 and ealier
papers of the author quoted therein

\bibitem{S2}  B. Schroer, \textit{Lightfront Formalism versus
Holography\&Chiral Scanning}, hep-th/0108203 and previous work cited therein

\bibitem{A-L}  C. D'Antoni and R. Longo, J. Funct. Anal. \textbf{15}, (1983)
199

\bibitem{Hal}  H. Halvorson, \textit{Reeh-Schlieder Defeats Newton-Wigner:
On alternative localization schemes in relativistic quantum field theories}
and references therein, quant-ph/0007060

\bibitem{Wald}  R. M. Wald, \textit{Quantum Field theory in Curved Spacetime
and Black Hole Thermodynamics}, University of Chicago Press 1994

\bibitem{Haag}  R. Haag, \textit{Local Quantum Physics}, Springer Verlag
(1992)

\bibitem{Requ}  M. Requardt, Commun. Math. Phys. \textbf{50}, (1979) 259

\bibitem{QED}  Selected papers on \textit{Quantum Electrodynamics}, edited
by J. Schwinger, Dover Publications Inc. New York 1958

\bibitem{Wa}  R. M. Wald, Phys.Rev. D48 (1993) 3427-3431

\bibitem{Bi-Wi}  J. J. Bisognano and E. H. Wichmann, J. Math. Phys. \textbf{%
17}, (1976) 303

\bibitem{Sewell}  G. L. Sewell, Ann. Phys. (N.Y.) \textbf{141}, (1982) 201

\bibitem{D-L}  S. Doplicher and R. Longo, Invent. math. \textbf{75}, (1984)
493

\bibitem{charges}  C. D'Antoni, S. Doplicher, K. Fredenhagen and R. Longo,
Commun. Math. Phys. \textbf{110}, (1987) 325

\bibitem{Sor}  M.Bombelli, R. K. Kaul, J Lee and R. D. Sorkin, Phys. Rev.D 
\textbf{34}, (1986) 373

\bibitem{trans}  B. Schroer, \textit{The Paradigm of a transverse Area
Structure for Lightfront Degrees of Freedom}, hep-th/0202085

\bibitem{BFV}  R. Brunetti, K. Fredenhagen and R. Verch, \textit{The
Generally Covariant Locality Principle}-\textit{A New Paradigm for Local
Quantum field Theory}, hep-th/01122000 and references on prior seminal work
of S. Hollands and R. M. Wald.

\bibitem{Lucio}  L. Fassarella and B. Schroer, \textit{The Fuzzy Analog of
Chiral Diffeomorphisms in higher dimensional Quantum Field Theories},
hep-th/0106064

\bibitem{Driessler}  W. Driessler, Acta Physica Austriaca \textbf{46},
(1977) 163, and references therein

\bibitem{Fard}  K. Ebrahimi-Fard, \textit{Comment on: Modular theory and
Geometry}, math-ph/001149

\bibitem{Joerss}  M. J\"{o}rss, Lett. Math. Phys. \textbf{38}, (1996) 252

\bibitem{Kosaki}  H. Kosaki, J. Operator Theory \textbf{16}, (1986) 335

\bibitem{Rehren}  K.-H. Rehren, Ann. Henri Poincar\'{e} \textbf{1}, (2000)
607

\bibitem{'t Hooft}  G. 't Hooft, in Salam-Festschrift, A. Ali et al. eds.,
World Scientific 1993, 284

\bibitem{Carlip}  S. Carlip, Nucl.Phys.Proc. Suppl.\textbf{88}, (2000) 10,
gr-qc/9912118

\bibitem{Wigner}  Lucio Fassarella and Bert Schroer, \textit{From Wigner
Particle Theory to Local Quantum Physics}, hep-th/

\bibitem{Wies}  H.-W. Wiesbrock, Comm. Math. Phys. \textbf{158}, (1993) 537

B. Schroer and H-W Wiesbrock, Rev. Math. Phys. \textbf{12}, (2000) 139

R. Kaehler and H.-W. Wiesbrock, JMP \textbf{42}, (2000) 74

\bibitem{Bu}  D. Buchholz, O. Dreyer, M. Florig and S. J. Summers, Rev.
Math. Phys. \textbf{12}, (2000) 475

\bibitem{S-W}  B. Schroer and H.-W. Wiesbrock Rev. Math. Phys. \textbf{12},
(2000) 461
\end{thebibliography}
\end{document}